\documentclass[a4paper]{jpconf}

\usepackage{citesort,iopams}
\usepackage{graphicx}

\newcommand{\bd}{\begin{displaymath}}
\newcommand{\ed}{\end{displaymath}}
\newcommand{\be}{\begin{equation}}
\newcommand{\ee}{\end{equation}}
\newcommand{\ba}{\begin{eqnarray}}
\newcommand{\ea}{\end{eqnarray}}

\begin{document}

\title{Particles, waves and trajectories:\\
210 years after Young's experiment}

\author{\'Angel S. Sanz}

\address{Instituto de F{\'\i}sica Fundamental (IFF--CSIC),
Serrano 123, 28006 - Madrid, Spain}

\ead{asanz@iff.csic.es}

\begin{abstract}
Mermin's ``shut up and calculate!''\ somehow summarizes the most widely
accepted view on quantum mechanics.
This conception has led to a rather constraining way to think and
understand the quantum world.
Nonetheless, a closer look at the principles and formal body of this
theory shows that, beyond longstanding prejudices, there is still room
enough for alternative tools.
This is the case, for example, of Bohmian mechanics.
As it is discussed here, there is nothing contradictory or wrong with
this hydrodynamical representation, which enhances the dynamical role
of the quantum phase to the detriment (to some extent) of the
probability density.
The possibility to describe the evolution of quantum systems in terms
of trajectories or streamlines is just a direct consequence of the fact
that Bohmian mechanics (quantum hydrodynamics) is just a way to recast
quantum mechanics in the more general language of the theory of
characteristics.
Misconceptions concerning Bohmian mechanics typically come from the
fact that many times it is taken out of context and considered as an
alternative theory to quantum mechanics, which is not the case.
On the contrary, an appropriate contextualization shows that Bohmian
mechanics constitutes a serious and useful representation of quantum
mechanics, at the same level as any other quantum picture, such as
Schr\"odinger's, Heisenberg's, Dirac's, or Feynman's, for instance.
To illustrate its versatility, two phenomena will be briefly
considered, namely dissipation and light interference.
\end{abstract}


\section{Introduction}
\label{sec1}

In 2013 we have celebrated the 100th anniversary of Bohr's atomic model
\cite{bohr100-1:2013,bohr100-2:2013}, which led to the development of
quantum mechanics in the 1920s.
However, we have also celebrated the 210th anniversary of Young's
famous two-slit experiment.
In 1803 Thomas Young presented to the Royal Society his experimental
``proof of the general law of the interference of two portions of
light'' \cite{young:PTRSL:1804,young:1807}; more than one hundred years
later this experiment has become one of the most influential ones in
physics \cite{shamos-bk}, particularly due to its tight connection with
quantum mechanics.
The fact is vividly expressed at the beginning of the third volume of
the {\it Feynman's Lectures on Physics} \cite{feynman:FLP3:1965}, where
we read:
\begin{quote}
In this chapter we shall tackle immediately the basic element of the
mysterious behavior in its most strange form.
We choose to examine a phenomenon which is impossible, {\it absolutely}
impossible, to explain in any classical way, and which has in it the
heart of quantum mechanics.
In reality, it contains the {\it only} mystery.
We cannot make the mystery go away by ``explaining'' how it works.
We will just {\it tell} you how it works.
In telling you how it works we will have told you about the basic
peculiarities of all quantum mechanics.
\end{quote}
The phenomenon referred to here, namely quantum interference, is
precisely the same phenomenon observed by Young ---although for light
instead of massive particles.

Nowadays the enormous success of quantum mechanics is indisputable.
Not only this theory explains the most fundamental aspects of the
physical world, but it has also given rise to technological
applications with a direct impact on our daily life.
Now, 210 years after Young's experiment, what do we really know about
quantum systems?
Unfortunately, not much (if anything at all).
That is, although we have a very accurate theory, our understanding of
this theory still relies on the ideas prevailing in the late 1920s and
1930s, strongly linked to the experimental capabilities at that moment.
At present fine experiments can be performed in the time domain,
reproducing Young's experiment particle by particle.
Even though the general conception of quantum systems is still anchored
in somewhat old-fashioned, self-imposed constraints, which have more to
do with positivist prejudices than with limitations of the theoretical
framework of quantum mechanics \cite{forman:HistStud:1971,jammer-bk:1966}.
Mermin's quotation ``shut up and calculate!''
\cite{mermin:PhysToday:1989,mermin:PhysToday:2004}
accurately summarizes this position.
Obviously, this has constituted (and still constitutes) an important
obstacle to the development and advance of new ways to understand the
quantum world, particularly those relying on the concept of trajectory.
This is the case, for example, of Bohmian mechanics.

The purpose of this communication is to show that there is nothing
contradictory or wrong with Bohmian mechanics.
Rather than a matter of taste, the discussion will show that this
approach is just another representation (a hydrodynamical one) of
quantum mechanics, at the same level as other more standard
representations, e.g., Schr\"odinger's, Heisenberg's, Dirac's,
Feynman's, etc.
Actually, although it goes beyond the scopes of this work, leaving
aside aspects commonly associated with Bohmian mechanics, like the
possibility of hidden variables or the ontology of the wave function,
it can readily seen that it is just a direct translation of quantum
mechanics into the more general language of the theory of
characteristics \cite{courant-hilbert-bk-2}.
Now, why Bohmian mechanics?
Because it focuses on the quantum phase, which is not a quantum
observable, although it plays a decisive role on quantum system
dynamics.
Bearing this in mind, this work is organized as follows.
The essential elements of this approach are introduced in
Sec.~\ref{sec2}.
In Sec.~\ref{sec3} a brief overview on how this approach and similar
ones have been applied to different physical problems is presented.
This makes readily apparent the nature of Bohmian mechanics as a
theory of characteristics, even if the latter's formalism is not
explicitly discussed.
In Sec.~\ref{sec4} Bohmian mechanics is applied to two different
phenomena, namely dissipation and light interference, in order to
show its versatility.
To conclude, in Sec.~\ref{sec5} a series of final remarks are
summarized.


\section{Waves, trajectories and quantum mechanics}
\label{sec2}

In appearance, the formulation of quantum mechanics in the
Schr\"odinger representation keeps a close analogy with classical
wave theory.
Within this representation, quantum systems are described by a
probability amplitude or wave function, $\Psi$, which evolves according
to the partial differential equation
\be
 i\hbar\ \! \frac{\partial \Psi}{\partial t} =
  -\frac{\hbar^2}{2m}\ \! \nabla^2 \Psi + V \Psi .
 \label{schro}
\ee
This equation describes the transport of probability instead of
energy, as it happens with classical waves.
Nonetheless, as Eq.~(\ref{schro}) is formulated, this is not evident,
since it rather displays the form of a typical diffusion equation
(with a complex diffusion coefficient).
After Max Born proposed the statistical interpretation, we say that
the probability density, $\rho \equiv |\Psi|^2$, gives the probability
that the quantum system has (or is in) a particular configuration.
Therefore, a representation of $\rho$ in time gives us information on
how the configurational probability of the system evolves (is
transported) in time, i.e., which are the configurations (e.g.,
positions) where we have more or less chance to find the system.
If we look at a typical outcome obtained from an interference
experiment, even in the case of very large and complex molecular
systems \cite{arndt:NNanotech:2012}, we find that these particles
distribute according to $\rho$, in agreement with Born's statistical
interpretation.
However, particles are detected one by one at different places.
The individual positions (evolutions) of these particles, however, is
not a quantum observable.

Motivated by that fact, in 1952 David Bohm proposed \cite{bohm:PR:1952-1}
a model that could explain such individual arrivals at the same time
that could also account for the collective particle behavior, all
without appealing to von Neumann's reduction postulate.
He regarded the individual systems as ``hidden variables''.
To develop this interpretative model, Bohm considered Schr\"odinger's
equation (\ref{schro}) plus the nonlinear (polar) transformation
\begin{equation}
 \Psi({\bf r},t) = \rho^{1/2}({\bf r},t) e^{iS({\bf r},t)/\hbar} .
 \label{e2}
\end{equation}
This transformation relates the complex-valued fields $(\Psi,\Psi^*)$
with the real-valued fields $(\rho,S)$, where $S$ describes the local
variation of the quantum phase.
Substituting (\ref{e2}) into the time-dependent Schr\"odinger equation
and then separating the real and imaginary parts of the resulting
equation, one finds
\setlength\arraycolsep{2pt}
\begin{eqnarray}
 \frac{\partial \rho}{\partial t} & + &
  \nabla \! \cdot \! \left( \rho\ \! \frac{\nabla S}{m} \right) = 0 ,
 \label{e3} \\
 \frac{\partial S}{\partial t} & + & \frac{(\nabla S)^2}{2m} + V
 - \frac{\hbar^2}{2m} \frac{\nabla^2 \rho^{1/2}}{\rho^{1/2}} = 0 .
 \label{e4}
\end{eqnarray}
The continuity (or conservation) equation (\ref{e3}) rules the ensemble
dynamics, i.e., the number of particles described by the probability
density has to remain constant; the quantum Hamilton-Jacobi
equation (\ref{e4}) describes the time-evolution of
the phase field, where the last term on the right-hand side is the
so-called quantum potential.
These two equations make more apparent the physical meaning of
Eq.~(\ref{schro}) as a transport equation.
In analogy to the classical Hamilton-Jacobi equation, if $S$
is identified with the classical action, then a momentum
${\bf p} = \nabla S$ can be postulated (Bohm's momentum).
This momentum can be expressed as ${\bf p} = m\dot{\bf r}$, which
gives rise to the guidance equation of motion
\begin{equation}
 \dot{\bf r} = \frac{\nabla S}{m} .
 \label{e10}
\end{equation}
The local velocity field ${\bf v} = \dot{\bf r}$, which depends
directly on the quantum phase, thus governs the individual system
dynamics and, in virtue of Eq.~(\ref{e3}), gives rise to averages
that are in agreement with the results obtained directly from
Schr\"odinger's equation.
It is interesting to note at this point that Eq.~(\ref{e10}) does
not need to be postulated, but it arises automatically within the
theory of characteristics \cite{courant-hilbert-bk-2}.
The integration of this equation in whichever evolution parameter
(e.g., time) generates the corresponding characteristics perpendicular
to $S$-surfaces of constant phase.

Previous to Bohm's model, de Broglie formulated a similar one
\cite{broglie:CompRend-2:1926} where particles were assumed to be
singularities guided by the wave.
Both models pursued essentially the same idea, namely to explain the
ensemble behavior of quantum systems at the same time that their
individual motion was also properly described.
The meaning of individual system, however, is rather uncomfortable,
since its (individual) evolution is not accessible.
To understand this idea with a simple example, think of the stream of
a river: its precise characterization does not provide any information
at all on the individual motion of its molecular constituents.
In this regard, it is probably more precise the approach developed by
Erwin Madelung in 1926.
Shortly after Schr\"odinger proposed his equation, Madelung provided a
clear prescription \cite{madelung:ZPhys:1926} to reformulate such an
equation in hydrodynamic form.
Accordingly, quantum systems could be visualized in terms of a series
of streamlines, which would follow the flow associated with the
system probability density.
That is, we know nothing about the precise motion of individual
systems (they could be evolving in a Brownian-like fashion,
for example), but still we can determine how the ensemble probability
flows throughout the corresponding configuration space.

Following Madelung's view, Eq.~(\ref{e10}) needs not be postulated,
but it is a consequence of the fact that Eq.~(\ref{e3}) is a
(real-valued) transport equation.
This allows us to establish a direct analogy with classical fluid
dynamics.
Hence, making use of the local probability current density,
\be
 {\bf J} = \frac{\hbar}{m}\ \! {\rm Im} \left(\Psi^* \nabla \Psi\right)
 = {\bf v} \rho ,
\ee
where ${\bf v}$ is a local (hydrodynamic)
velocity field, one finds the expression
\begin{equation}
 {\bf v} = \frac{{\bf J}}{\rho} = \frac{\nabla S}{m} ,
 \label{e10b}
\end{equation}
which is equivalent to the above Bohmian velocity.
The integration in time (or whichever parameter) of Eq.~(\ref{e10b})
generates a family of streamlines or paths (for each wave function
considered) along which the quantum fluid propagates, just as in the
case of a classical fluid.
As it can be inferred, the first equality goes beyond Bohmian mechanics
and allows to define streamlines in any system characterized by
a certain density and a vector that transports it through the
corresponding configuration space, regardless whether such a
density describes a quantum system or not.
Notice that in this case, instead of the set of equations (\ref{e3})
and (\ref{e4}), we have a set of hydrodynamic equations, as shown
by Takabayasi \cite{takabayasi:ProgTheorPhys:1952,takabayasi:ProgTheorPhys:1953,takabayasi:ProgTheorPhys:1983}.

\begin{figure}[t]
\centering
  \includegraphics[width=11cm]{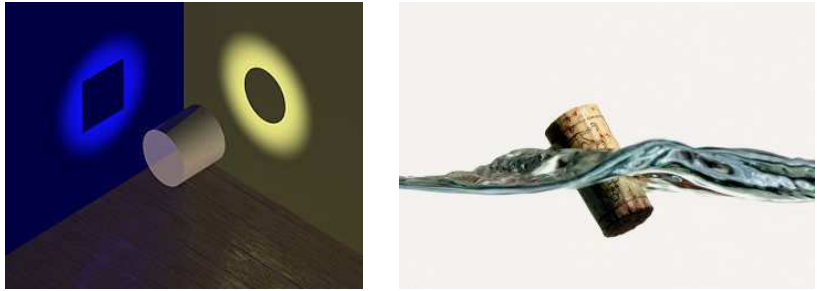}
  \caption{\label{fig1}
   The different representations of quantum mechanics provide us with a
   different description of quantum systems.
   Typically quantum systems are thought from a complementary viewpoint
   (left).
   Bohmian mechanics, however, gives preference to the configuration
   space and emphasizes transport-related properties (right).}
\end{figure}

Taking into account the fact that Bohmian mechanics is a different
theory, but only a reformulation of quantum mechanics, one may wonder
why it is worth using.
The answer is very simple, it is just a matter of which aspect of the
physical system one wishes to stress.
This is nicely illustrated in Fig.~\ref{fig1}.
Typically, we understand quantum systems in a somewhat dichotomic
way, e.g., either wave or particle, position or momentum, energy or
time, etc, emphasizing the so-called complementarity (see left panel).
However, we may also be interested in transport issues, for which
Bohmian mechanics constitutes a more convenient picture (right panel),
since it describes the probability flow in configuration space
without caring about any other complementary aspect.
In this regard, we find, for example, that probably one of the most
relevant and distinctive properties revealed by this quantum
hydrodynamic formulation, is the non-crossing
\cite{sanz:JPA:2008,sanz-bk-1}, i.e., the fact that
quantum fluxes (described in terms of bunches of Bohmian trajectories)
cannot cross in configuration space at the same time.
This is in compliance with the outcomes obtained from a recent
photon-by-photon realization of Young's experiment
\cite{kocsis:Science:2011}.
As a direct consequence of the non-crossing, one is able to establish
well-define quantum probability tubes \cite{sanz:JPA:2011,sanz:AnnPhys:2013}, i.e.,
tubes in configuration space along which the integral of the
probability density at a given time remains constant.


\section{Hydrodynamic approaches in the literature\footnote{The
references here provided are rather scarce, but I think that they will
be enough for the reader interested in further enquiring about the
use of streamlines/trajectories in the literature.}}
\label{sec3}

As seen above, the hydrodynamic language of Bohmian mechanics enables a
visualization of quantum systems in terms of streamlines that follow
the flow of the probability density.
This simple pictorial representation attracted the attention of the
chemical physics community immediately after the first wave-packet
propagation schemes were developed and used in the field by the end
of the 1960s.
Pioneers in this field were McCullough and Wyatt, who studied the
quantum dynamics of collinear atom-diatom reactions in hydrodynamic
terms \cite{mccullough:JCP:1969,mccullough:JCP:1971-1,mccullough:JCP:1971-2},
just in a period when classical trajectories were in fashion.
The following quotation, extracted from \cite{mccullough:JCP:1971-1},
gives a very clear idea of what this community was looking for:
\begin{quote}
Classical mechanics gives an amazingly good description of the
probability density and flux patterns during most of the reaction;
however, the classical and quantal descriptions begin to diverge near
the end of the reaction.
Essentially, the classical reaction terminates before the quantal
reaction.
The dynamic behavior of the reaction is hydrodynamically turbulent, as
shown by transient whirlpool formation on the inside of the reaction
path.
\end{quote}
This ended up with the development of the so-called quantum-trajectory
methods \cite{wyatt-bk} at the end of the 1990s, just something that
David Bohm himself thought to be the path to follow to describe the
physical systems.

In above case, a hydrodynamical viewpoint was adopted, without making
an explicit calculation of streamlines or trajectories (hydrodynamical
vector fields were used instead).
Nevertheless, a few years later, in the middle of the 1970s,
Hirschfelder and coworkers materialized Madelung's ideas when trying
to recast solutions of Schr\"odinger's equation in a pictorial way
for tunneling and vortical dynamics.
As before, the underlying motivation was seeking for a better
understanding of the phenomena studied \cite{hirsch:JCP:1974-1}:
\begin{quote}
This paper has resulted from an effort to get a better understanding of
quantum mechanics by making a thorough study of a very simple problem,
[\ldots].\
The mathematics is simple, but the analysis is far-reaching.
\end{quote}
Again, these studies came from the field of the chemical physics.
In this sense, it is interesting to note that the first studies of
barrier tunneling came from this community \cite{hirsch:JCP:1974-1},
thus predating those that later on appeared following Bohm's ideas
\cite{hiley:foundphys:1982}.

In the 1980s we find a convergence to Bohmian mechanics from the field
of molecular magnetism \cite{lazzeretti:IJQC:1984,lazzeretti:IJQC:2011},
where the quantum hydrodynamics was also in fashion.
In particular, in \cite{lazzeretti:IJQC:1984} we read:
\begin{quote}
A representation of the electron flow induced by the external field can
be extremely useful to understand molecular magnetism.
To this end, maps reporting modulus and trajectory of quantum-mechanical
current density revealed a fundamental tool, whose importance could be
hardly overestimated.
\end{quote}
By means of these representations one can observe very convolved
trajectories describing the electron current densities induced by
external magnetic fields acting on different types of molecular
systems \cite{lazzeretti:IJQC:2011}.
At an applied level, ``these tools provide fundamental help for
rationalization of magnetic response properties, such as magnetizability
and nuclear magnetic shielding'' \cite{lazzeretti:IJQC:2011}.

The same motivation can be found even earlier in electromagnetism,
where the tradition of explaining phenomena in terms of rays was
stronger and therefore there was not an urgent need to appeal to
the Madelung-Bohm scheme.
Thus, in the 1952, the same year that Bohm published his work on
hidden variables, Braunbek and Laukien published \cite{braunbek:Optik:1952}
a work where they studied the diffraction by an edge (a perfectly
conducting half-plane) by means of lines of average electromagnetic
energy flow, obtained from the analytical solution provided to this
problem by Sommerfeld in 1896 \cite{sommerfeld:MathAnn:1896}.
About 20 years later (again in the 1970s, when the computational tools
were already more sophisticated), Prosser produced
\cite{prosser:ijtp:1976-1} the first trajectories
for Young's two slit experiment and provided an explanation in terms
of ``photon'' trajectories \cite{prosser:ijtp:1976-2}.
These trajectories preceded those obtained by Dewdney \etal for matter
waves using Bohm's model \cite{dewdney:NuovoCimB:1979}.
Later on, different authors have treated the problem of electromagnetism
in terms of trajectories (an account can be found in \cite{sanz:AnnPhysPhoton:2010}),
until in 2011 Steinberg and coworkers performed an experiment from which
the first trajectories were inferred experimentally \cite{kocsis:Science:2011}.
At a more applied level, and in consonance with Madelung's viewpoint,
we also find works dealing with streamlines in wave optics
\cite{zakowicz:PRE:2001,sanz:JOSAA:2012} or transport through
billiards \cite{sadreev:PRE:2004}, for example.

Of course, one could regard the trajectories for electromagnetic fields
(radiation) as similar to those for matter waves (massive particles).
In the end, Maxwell's equations can be seen as equivalent to
Schr\"odinger's one \cite{scully-zubairy-bk}.
However, we can move apart from these scenarios, and still we find
analogous streamline-based descriptions with similar purposes.
This is the case sound waves, for example, some of which started
appearing in the middle 1980s
\cite{waterhouse:JASA:1985,waterhouse:JASA:1986-1,waterhouse:JASA:1986-2,waterhouse:JASA:1987-1,waterhouse:JASA:1987-2}.
In particular, in Ref.~\cite{waterhouse:JASA:1985} we read:
\begin{quote}
A method is presented for computing the energy streamlines of a sound
source.
This enables charts to be plotted showing, as continuous lines, the
flow paths of the sound energy from the vibrating surface to the
nearfield and beyond.
Energy streamlines appear to be a new construct; they have some
similarities to the velocity streamlines used in fluid dynamics.
Examples of the energy streamlines are given for the point-driven
plate in water.
[\ldots] These streamlines make it easer for the eye to follow the
energy flow from the source into the nearfield and beyond.
These paths are complicated, in some cases, but are of considerable
interest from several points of view.
\end{quote}
This is precisely the role played by Bohmian mechanics in the study of
quantum mechanical systems!
This role is clearer and clearer as one goes through
Ref.~\cite{waterhouse:JASA:1985} in more detail.
While searching through the literature, one finds remarkable the fact
that in other areas of physics dealing with waves (other than quantum
mechanics or, by extension, quantum optics) the visualization of flows
is particularly relevant.
Physicists have tried to develop methods based on characteristics, like
those described above, or just stroboscopic interferometric ones
\cite{chevalerias:JOSA:1957}, in order to explore and understand the
behavior of the fluid qualitative and quantitatively.
This is in contrast with the reluctancy found in quantum mechanics to
treat systems on equal footing.

According to the above discussion, the interest generated
by the calculation of streamlines associated with the transport of some
quantity ---e.g., probability, electromagnetic energy, pressure energy---
has led to a series of approaches that all converge to the very same
need: dealing with a tool that allows one to objectively monitor the
transport of such a quantity.
In this regard, if quantum mechanics is separated from any other wave
theory, one may end up concluding that Bohmian trajectories represent
real trajectories pursued by real particles (or, in general, degrees
of freedom), i.e., they are hidden variables.
However, if this theory is properly contextualized, we find that there
is nothing that allows us to establish a direct connection between the
possible motion of real particles and this type of trajectories
\cite{sanz:AJP:2012,sanz-bk-1}.
Comparing with other theories, and as it will be seen in the next
Section, as much we can say that a Bohmian particle is a particle that
obeys a Bohmian dynamics, i.e., according to a (local) average drift
momentum, which provides the particle with nonlocal (global)
hydrodynamic-like information.
Such a particle allows us to infer dynamical properties of the quantum
fluid, which are usually ``hidden'' when studied by means of the wave
function formulation.
That is, Bohmian particles are the quantum equivalent of classical
tracer particles (or just tracers) that can be found in other areas of
physics and chemistry.


\section{Some applications}
\label{sec4}


\subsection{Dissipative Bohmian mechanics}
\label{sec41}

The first context where we are going to apply Bohmian mechanics in the
sense described above, namely as a hydrodynamic picture of quantum
mechanics, is that of dissipation.
Thus, consider the well-known classical dissipative equation
\be
 m\ddot{x} + m\gamma\dot{x} + \frac{\partial V(x)}{\partial x} = 0 ,
 \label{dis1}
\ee
where $m$ is the system mass and $\gamma$ is the friction coefficient.
According to this equation, the dissipation undergone by the system is
proportional to its speed at a given time, $\dot{x}$.
In order to determine the quantum analog, i.e., to specify the
Schr\"odinger equation corresponding to Eq.~(\ref{dis1}), we need
to find a suitable Hamiltonian.
One way to proceed is by considering \cite{sanz-bk-2} the change of
variables
\be
 X = x , \qquad P = m e^{\gamma t} \dot{x} = p e^{\gamma t} ,
 \label{dis3}
\ee
where $(x,p)$ denote the physical variables, with $p=m\dot{x}$, and
$(X,P)$ the canonical ones.
The latter allow us to define the conservative Hamiltonian
\be
 \mathcal{H}_{(X,P)} = \dot{X}P - \mathcal{L}
  = \frac{P^2}{2m}\ \! e^{-\gamma t} + V(X) e^{\gamma t} ,
 \label{dis4}
\ee
and satisfy the usual canonical relations
\be
 \dot{X} = \frac{\partial \mathcal{H}_{(X,P)}}{\partial P} , \qquad
 \dot{P} = - \frac{\partial \mathcal{H}_{(X,P)}}{\partial X} ,
 \label{dis6}
\ee
which enable the conservation of the total energy.
If the initial energy is $E_0$ and therefore $\mathcal{H}_{(X,P)}=E_0$,
the inverse change to the physical coordinates gives
\be
 \mathcal{H}_{(x,p)} = \frac{p^2}{2m} + V(x) = E_0 e^{-\gamma t} ,
 \label{dis7}
\ee
i.e., the energy is lost exponentially at a constant rate ($\gamma$),
as expected from the physical dissipative system described by
Eq.~(\ref{dis1}).
The dissipative model given by Eq.~(\ref{dis4}) is known as the
Caldirola-Kanai model \cite{caldirola:NuovoCim:1941,kanai:ProgTheorPhys:1948}
and constitutes one of the former attempts
to express (\ref{dis1}) in a Hamiltonian form.

In order to find the quantum analog of the Caldirola-Kanai model, we
now make use of the standard quantization procedure, and associate the
operators $\hat{X}$ and $\hat{P} = - i\hbar \partial/\partial \hat{X}$
with the canonical variables $X$ and $P$, respectively.
Because of their canonicity, these operators satisfy the usual
commutation relation $[\hat{X},\hat{P}] = i\hbar$ ---which does not
hold for the ``physical'' operators, $[\hat{x},\hat{p}] = i\hbar
e^{-\gamma t}$.
Accordingly, the quantum Caldirola-Kanai Hamiltonian arises by
replacing $X$ and $P$ in Eq.~(\ref{dis4}) by the
corresponding operators,
\be
 \hat{\mathcal{H}}_{(\hat{X},\hat{P})} =
  -\frac{\hbar^2}{2m}\ \! e^{-\gamma t}
   \frac{\partial^2 \phantom{\Psi}}{\partial X^2}
  + e^{\gamma t} \hat{V}(\hat{X}) ,
 \label{dis11}
\ee
and the corresponding Schr\"odinger equation, in the physical variable
$x$, reads as
\be
 i\hbar\ \! \frac{\partial\Psi}{\partial t} =
  - \frac{\hbar^2}{2m}\ \! e^{-\gamma t}\ \!
     \frac{\partial^2\Psi}{\partial x^2}
  + e^{\gamma t} V(x) \Psi .
 \label{dis12}
\ee
It is easy to show that the associated (dissipative) Bohmian
trajectories obey the modified guidance equation
\be
 \dot{x} = \frac{J}{\rho}
  = \frac{1}{m} \frac{\partial S}{\partial x}\ \! e^{-\gamma t} .
 \label{dis13}
\ee
Note the similarity between this equation and the classical analog,
$\dot{x} = p e^{-\gamma t}/m$.

\begin{figure}[t]
\centering
 \includegraphics[width=16cm]{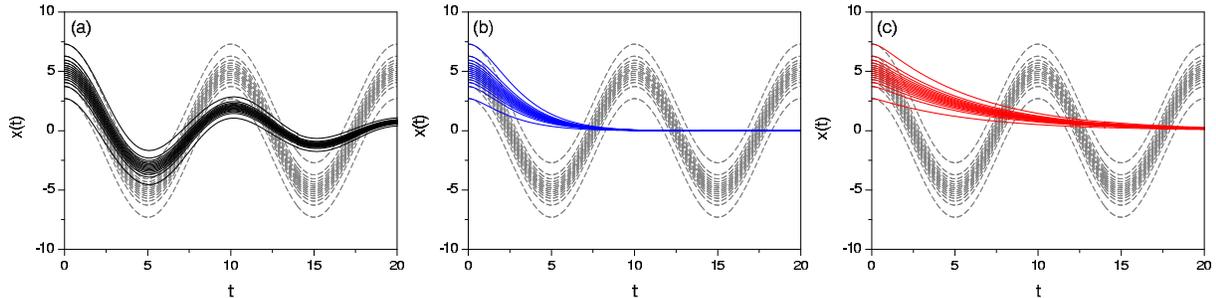}
  \caption{\label{fig2}
   Dissipative Bohmian trajectories for: (a) $\gamma = 0.3\omega_0$,
   (b) $\gamma = 2\omega_0$, and (c) $\gamma = 4\omega_0$, with
   $\omega_0 = 2\pi/\tau_0 \approx 0.628$ ($\tau_0 = 10$).
   To compare with, the frictionless Bohmian trajectories have also
   been included in each panel (gray dashed lines).
   The initial positions have been distributed according to the initial
   Gaussian probability density.}
\end{figure}

\begin{figure}[t]
\centering
 \includegraphics[width=16cm]{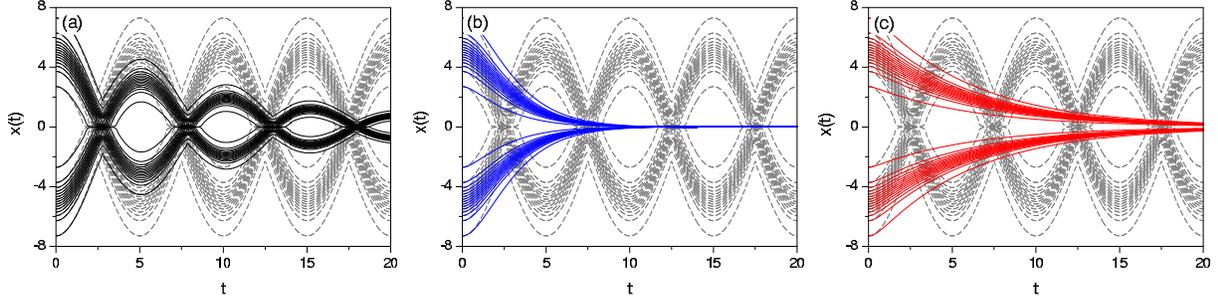}
  \caption{\label{fig3}
   Dissipative Bohmian trajectories for: (a) $\gamma = 0.3\omega_0$,
   (b) $\gamma = 2\omega_0$, and (c) $\gamma = 4\omega_0$, with
   $\omega_0 = 2\pi/\tau_0 \approx 0.628$ ($\tau_0 = 10$).
   To compare with, the frictionless Bohmian trajectories have also
   been included in each panel (gray dashed lines).
   The initial positions have been distributed according to the initial
   Gaussian probability density.}
\end{figure}

For potential functions $V(x)$ which are polynomials with a degree
equal or smaller than two, one can find analytical solutions for
initial wave functions with the shape of a Gaussian wave packet
\cite{sanz-bk-2,sanz:arxiv:dissip}.
Without entering details, just to illustrate the dissipative dynamics
described by Eq.~(\ref{dis12}), we are going to consider a harmonic
oscillator.
The dissipative Bohmian trajectories for different values of the
friction coefficient are displayed in Fig.~\ref{fig2}.
As it can be noticed, in the physical coordinates not only there is a
clear violation of the usual commutation relation, which manifests in
a vanishing dispersion of the wave packet, but the system approaches
the bottom part of the potential well, thus going down the zero-point
energy.
This is a pathological behavior (in the physical variables) typical of
the Caldirola-Kanai model, which is based on a continuous dissipation
of the system energy due to the lack of a proper quantization.
The latter only applies to the canonical variables $X$ and $P$.
A similar behavior can be found, for example, when dealing with
beables \cite{vink:PRA:1993,lorenzen:PRA:2009}.
The same correspondence can also be found if one considers an initial
wave-packet superposition inside the harmonic potential, as shown
in Fig.~\ref{fig3}.

In order to avoid such an inconvenience, other models have been
proposed in the literature as suitable quantum candidates of
Eq.~(\ref{dis1}) \cite{hasse:JMathPhys:1975,schuch:PRA:1983,schuch:PRA:1984-1,schuch:PRA:1984-2,schuch:PRA:1997}.
In analogy to Brownian-type wave-function models \cite{sanz:CJC:2014},
these models use to include a nonlinear term, thus also making the
corresponding Schr\"odinger equation to be nonlinear.
Of course, this nonlinear term is usually accompanied by a stochastic
term that accounts for the random fluctuations of the medium that give
rise to such a nonlinearity (just as in the Langevin equation for
Brownian motion, for example).
For example, in Kostin's model \cite{kostin:jcp:1972} a nonlinear term
depending on the phase of the wave function is considered, with the
Schr\"odinger equation reading as
\be
 i\hbar\ \! \frac{\partial\Psi}{\partial t} =
  - \frac{\hbar^2}{2m} \frac{\partial\Psi}{\partial x^2}
  + V \Psi + V_R \Psi
  + \gamma \left( S - \int \!\! \rho \ \! S dx \right) \Psi ,
 \label{nonschro}
\ee
with $S = (\hbar/2i) \ln (\Psi/\Psi^*)$ and where $V_R$ is a random
potential.
Here we have focused in this model in particular because of the link
that can be established between this model and Bohmian mechanics
precisely through $S$, as it will be seen below.

In the full dissipative case, where $V_R = 0$,
if we substitute the usual polar ansatz into Eq.~(\ref{nonschro}) and
then proceed as in standard Bohmian mechanics, we reach the modified
quantum Hamilton-Jacobi equation
\be
 \frac{\partial S}{\partial t}
 + \frac{1}{2m}\! \left(\frac{\partial S}{\partial x}\right)^2
 + V + Q + \gamma S = 0,
 \label{kostin2}
\ee
where $Q$ denotes the quantum potential (see Sec.~\ref{sec2}).
Assuming that the Bohmian trajectories are obtained from the usual
equation of motion, $p = m\dot{x} = \partial S/\partial x$, and
differentiating Eq.~(\ref{kostin2}) with respect to $x$, we reach
\be
 m\ddot{x} + m\gamma\dot{x} + \frac{\partial (V+Q)}{\partial x} = 0 ,
 \label{kostin3}
\ee
where we have made use of the Lagrangian derivative
\be
 \frac{d}{dt}
 = \frac{\partial}{\partial t}
   + \dot{x}\ \! \frac{\partial}{\partial x}
 = \frac{\partial}{\partial t}
   + \frac{1}{m}\! \left(\frac{\partial S}{\partial x}\right)\!
     \frac{\partial}{\partial x} .
\ee
Note that Eq.~(\ref{kostin3}) has the same functional form as
Eq.~(\ref{dis1}), except for the fact that it also includes a purely
quantum force (given in terms of the space derivative of the quantum
potential).
That is, Bohmian mechanics shows a direct path to find the quantum
analog of Eq.~(\ref{dis1}), consisting of just including the quantum
potential.
Kostin's model has been proven very useful to determine energy bound
states due to its convergence by dissipation \cite{garashchuk:JCP:2013}.

\subsection{``Photon'' trajectories}
\label{sec42}

Another field of interest concerning the applications of Bohmian
mechanics is electromagnetism, as mentioned above, where a
time-independent trajectory approach can be readily developed starting
from Maxwell's equations \cite{sanz:PhysScrPhoton:2009,sanz:AnnPhysPhoton:2010}.
Within this approach, where the time-dependence is removed by averaging
over the oscillating electromagnetic field, the trajectories describe
the flow of electromagnetic energy (analogous to the probability
accounted for by the Bohmian trajectories).
As shown by Prosser \cite{prosser:ijtp:1976-1}, this is
possible by directly considering Maxwell's equations defined by an
electric field ${\bf E}({\bf r})$ and a magnetic field
${\bf H}({\bf r})$, the two key elements necessary to define the
corresponding streamlines are the time-averaged electromagnetic energy
density and Poynting vector (electromagnetic current density),
\begin{eqnarray}
 U({\bf r}) & = & \frac{1}{4}
  \left[ \epsilon_0 {\bf E}({\bf r}) \cdot {\bf E}^*({\bf r})
   + \mu_0 {\bf H}({\bf r}) \cdot {\bf H}^*({\bf r}) \right] .
 \label{e30} \\
 {\bf S}({\bf r}) & = & \frac{1}{2}\
  {\rm Re} \left[ {\bf E}({\bf r}) \times {\bf H}^*({\bf r}) \right] ,
 \label{e29}
\end{eqnarray}
respectively; ${\bf E}({\bf r})$ and ${\bf H}({\bf r})$ denote
respectively the spatial part of the electric and
magnetic field vectors, which satisfy Maxwell's equations and have
been assumed to be harmonic, i.e.,
\be
 \begin{array}{c}
  \tilde{\bf E}({\bf r}) = {\bf E}({\bf r}) e^{-i\omega t} , \\
  \tilde{\bf H}({\bf r}) = {\bf H}({\bf r}) e^{-i\omega t} .
 \end{array}
 \label{eq6}
\ee
Since the electromagnetic energy density is transported through space
in the form of the Poynting vector, a local velocity field can be
defined \cite{bornwolf-bk} in analogy to (\ref{e10b}), which reads as
\begin{equation}
 {\bf S}({\bf r}) = U({\bf r}) {\bf v} .
\end{equation}
The electromagnetic energy flow lines or ``photon'' paths are then
obtained by integrating the equation
\begin{equation}
 \frac{d{\bf r}}{ds} = \frac{1}{c}
  \frac{{\bf S}({\bf r})}{U({\bf r})}
\end{equation}
along the arc-length coordinate $s$ (which can be referred to a
proper time $\tau = s/c$, with $c$ being the speed of light).
The analogy between light and massive particles becomes more apparent
if the spatial parts of the electric and magnetic fields are expressed
in terms of a scalar function $\Psi$ that
satisfies the Helmholtz equation and the corresponding boundary
conditions \cite{sanz:PhysScrPhoton:2009,sanz:AnnPhysPhoton:2010}.

In order to show such an analogy, consider a monochromatic
electromagnetic wave in vacuum incident onto a plate with two slits.
The plate is on the $XY$ plane, at $z=0$.
For simplicity, it is assumed that the fields are independent of the
$y$ coordinate.
This assumption is well justified if the slits are parallel to the $y$
axis and their width along this axis is much larger than along $x$.
In such a case from Maxwell's equations one obtains two independent
sets of equations.
One involves the components $H_x$, $H_z$, and $E_y$ of the
electromagnetic field and is commonly regarded as $E$-polarized.
The other involves the components $E_x$, $E_z$, and $H_y$, and is
known as $H$-polarization.
The electric and magnetic fields behind the slits can be expressed as
\cite{sanz:PhysScrPhoton:2009}
\ba
 {\bf E}({\bf r}) & = &
   -\frac{i\beta}{k}\frac{\partial \Psi}{\partial z}\ \! {\bf e}_x
   +\frac{i\beta}{k}\frac{\partial \Psi}{\partial x}\ \! {\bf e}_z
   +\alpha \Psi {\bf e}_y ,
 \label{eq9} \\
 {\bf H}({\bf r}) & = &
    \frac{i\alpha}{\omega\mu_0}\frac{\partial \Psi}{\partial z}\ \!
     {\bf e}_x
   -\frac{i\alpha}{\omega\mu_0}\frac{\partial \Psi}{\partial x}\ \!
     {\bf e}_z
   +\frac{k\beta e^{i\varphi}}{\omega\mu_0}\ \! \Psi {\bf e}_y .
 \label{eq10}
\ea
The scalar field $\Psi$ at any $z$ can be expressed as a
Fresnel-Kirchhoff integral \cite{dimic:PhysScr:2013}, which in general
has to be numerically integrated once the initial condition is
established.
In particular, if one considers a grating with two Gaussian slits
\cite{feynman-bk2}, the initial condition can be assumed to be a
coherent superposition of the two diffracted waves,
\be
 \Psi(x,0) = \psi_1(x,0) + \psi_2(x,0) ,
 \label{eq12}
\ee
where
\be
 \psi_i(x,0) = \left(\frac{1}{2\pi\sigma_i^2}\right)^{1/4}
  e^{-(x-x_{0,i})^2/4\sigma_i^2} W(x-x_{0,i},w_i) ,
 \label{eq13}
\ee
with $i=1,2$.
A window function, $W(x,w)$, has been added to each wave packet in
order to provoke a truncation and, therefore, to analyze the eventual
effects on the final interference pattern.
In this case, this function is such that it is one within the extension
covered by the corresponding slit (i.e., between $-w_i$ and $w_i$),
and zero everywhere else.

\begin{figure}
 \begin{center}
 \includegraphics[width=14cm]{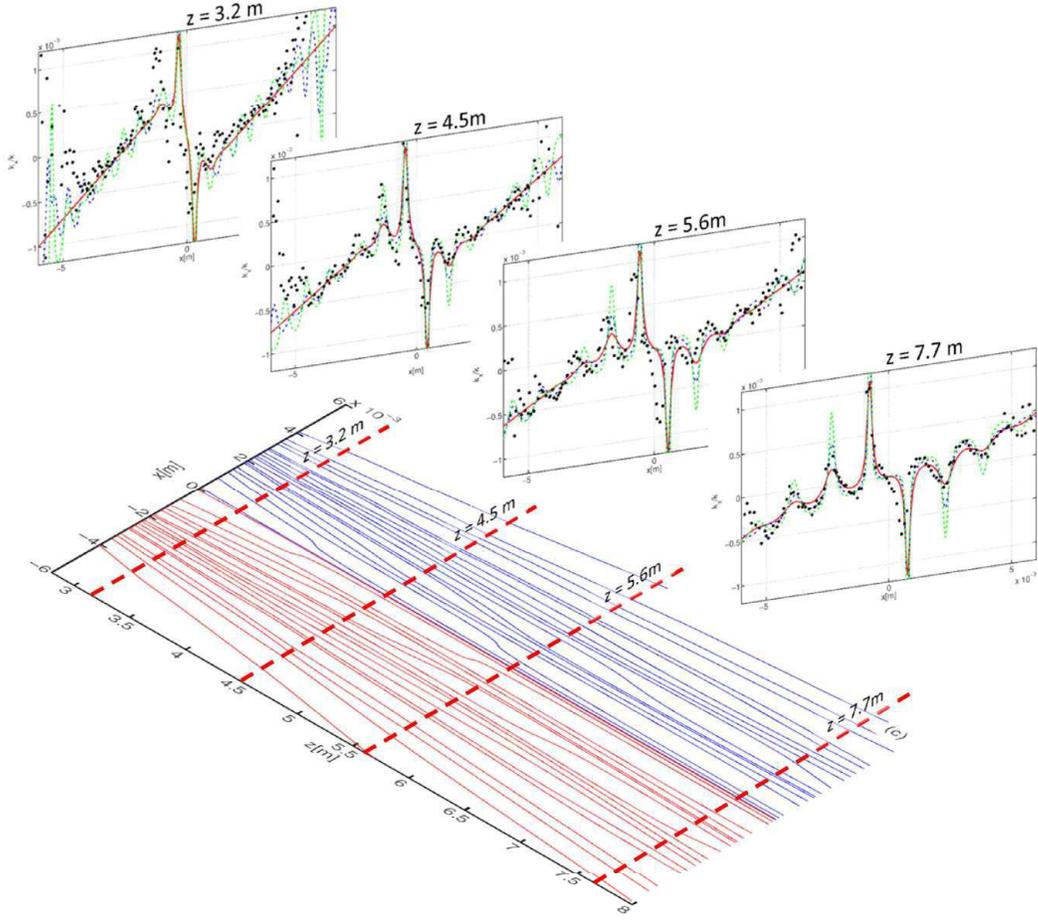}
 \caption{\label{fig4}
  Lower panel: Electromagnetic energy streamlines or average ``photon''
  trajectories behind two Gaussian slits \cite{feynman-bk2}.
  The numerical parameters used are \cite{dimic:PhysScr:2013}:
  $\sigma_1 = \sigma_2 = 0.3$~mm, $x_{0,1} = -x_{0,2} = 2.35$~mm, and
  $\lambda = 943$~nm.
  The trajectories have been distributed according to the corresponding
  initial Gaussian probability densities behind each slit.
  Upper panels: Transverse momentum, $k_x/k$, evaluated at the four
  distances reported in Ref.~\cite{kocsis:Science:2011}
  (see red dashed lines in the lower panel) for full Gaussian slits
  (red solid lines) and slits truncated at $w_i = 1.9\sigma_i$ (blue
  dotted lines) and $w_i = 1.5\sigma_i$ (green dashed lines).
  To compare with, the experimental data (black circles) are also
  displayed.
  In these four calculations, asymmetric Gaussians were used, with
  parameters: $\sigma_1 = 0.307$~mm, $\sigma_2 = 0.301$~mm,
  $x_{0,1} = 2.335$~mm, $x_{0,2} = -2.355$~mm, and $\lambda = 943$~nm.}
 \end{center}
\end{figure}

A series of electromagnetic energy flow lines or averaged ``photon''
trajectories are displayed in Fig.~\ref{fig4} (upper panel).
As in the previous example, the initial positions have also been
distributed according to the Gaussian weight.
To compare with, the numerical values considered have been chosen as in
the experiment reported in Ref.~\cite{kocsis:Science:2011}.
Although the model is rather simple and makes no explicit use of the
concept of ``weak measurement'', there is a good agreement between the
simulation and the trajectories inferred from the experimental data,
which shows the suitability of Maxwell's equations to describe this
type of experiments.
This adequacy is also seen when the transversal momentum obtained from
the Poynting vector is compared to the corresponding experimental data,
as shown in the upper panels for four different distances from the two
slits.
Notice that these two quantities are connected through the relation
\cite{barnett:PhilTRSA:2010}
\be
 \frac{k_x}{k} = \frac{S_x}{S} ,
 \label{eq17}
\ee
where $S_x$ and $S$ refer to the $x$-component and modulus of the
Poynting vector, respectively.
In particular, in the four upper panels displayed in Fig.~\ref{fig4}
the Gaussians have been considered to be asymmetric in order to find a
better fitting with the experimental data (black circles).
It can be seen that as the Gaussians are more severely truncated, the
oscillations of $k_x/k$ undergo a remarkable increase.
This means a stronger action of the quantum potential between adjacent
interference fringes according to the usual Bohmian interpretation
\cite{dewdney:NuovoCimB:1979,sanz:JPCM:2002}.


\section{Final remarks}
\label{sec5}

In general, the position maintained with respect to Bohmian mechanics
is that it constitutes an alternative interpretation to quantum
mechanics (or even an alternative theory).
From the above discussion, it is difficult to find arguments sustaining
such claims ---often used in a pejorative sense to refute any work in
this area.
We have seen that Bohmian mechanics should be rather regarded as an
alternative and complementary representation of quantum mechanics,
particularly when we note that it is the direct translation of the
latter into the language of the theory of characteristics, which has
nothing to do with hidden variables or ontological views.

In that sense, it is worth thinking for a while about the different
representations that we have in classical and quantum mechanics.
For the former, we admit different formulations, each one emphasizing
a different physical aspect of the systems described.
For example, Newton's formulation relies on the relationship between
the motion displayed by objects and the external forces that act on
them.
On the other hand, we also have Hamilton's formulation, which is based
on the concept of energy conservation.
This allows us to tackle physical problems in a rather flexible way,
choosing the formulation that better fits our needs, and at the same
time to understand the same phenomenon from different perspectives.

Similarly in quantum mechanics there are also different formulations
or ways to tackle the same problem, which are chosen according to their
suitability ---either analytical or numerical.
For example, Schr\"odinger's formulation stresses the time-evolution of
the system (through its wave function) under the influence of a given
Hamiltonian; any property of the system is synthesized from its wave
function either at a given time (probability densities) or at two
different times (correlation functions).
With Heisenberg's formulation, on the contrary, one focuses on the
evolution in time of the operators associated with observables and
how they act on a given (time-independent) system state.
Half-way between both, the Dirac or interaction representation is more
convenient to to analyze the dynamics of systems interacting with other
systems.
Feynman's path-integral representation, relying on the concept of
classical action and trajectory, is suitable in the treatment of large
systems.
Other representations, such as the Wigner-Moyal or the Husimi ones,
stress the role of the density matrix in phase space.
All these well-known examples of equivalent formulations of quantum
mechanics, which provide a different description of the same system.
In the same way, Bohmian mechanics stresses the role of the quantum
phase, which has determining consequences for quantum systems even
though it is not an observable.
Figure~\ref{fig1} summarizes this descriptive complementarity.

The trajectories that we obtain through Bohmian mechanics help us to
visualize and understand the physics underlying quantum systems and
phenomena by monitoring the flow of the probability.
This appealing feature has been used in different contexts, not
necessarily connected to Bohmian mechanics or, in general, to quantum
mechanics.
Here, for example, the dynamics of a dissipative system has been
analyzed, making evident the pathologies of a well-known model and
how they can be avoided.
In this sense, noticed that at a practical level the calculation of
Bohmian trajectories is more convenient even if the Schr\"odinger
equation has to be solved in order to compute them.
First, the analysis of the system dynamics in terms of trajectories
is simpler than in terms of probability densities, specially for two
or more dimensions.
Second, if instead of probabilities, one considers expectation values,
the trajectories are still very appropriate, because the former only
provide us with averaged information at every time, while the latter
allow us to visualize the dynamics of each particular part of the
quantum state.

Furthermore, the same concepts and tools can be extended to other areas
of physics, as we have seen in the example of the ``photon''
trajectories.
Although the differential equations are different and the fields
propagated have a different physical meaning, the underlying ideas
are exactly the same.
This is actually the way how the recent experimental realization of
Young's two slit experiment has been explained \cite{kocsis:Science:2011}.
Among the many different ways that one could devise to join the
transversal momentum function at different distance from the two
slits, Bohmian mechanics provides a very clear prescription of how
to do it, without incurring any kind of approximation or including
any external element to quantum mechanics.

In that latter regard, I shall finish here with the following
consideration.
In order to explain his outcomes from the two-slit experiment, Young
used the so-called Huygens' construction.
Accordingly, the position of a wavefront at a given time can be
obtained by considering a wavefront at a previous time.
Each point on this second wavefront is assumed to be a source of
secondary waves, whose interference gives rise to the wavefront at
the later time.
The direction of travel of these wavefronts is what we call a ray.
In the case of plane waves, the wavefronts are perpendicular to rays;
in the case of interference, the shape of the wavefronts varies from
point to point, that making intractable a ray description \ldots from
an analytical viewpoint!
Should Young have a computer, he would have been able to evaluate the
rays locally, at each point \ldots and would have discovered
Bohmian mechanics more than 100 years earlier, for Bohmian trajectories
constitute the convoluted generalization of the concept of ray.
In other words, they are the characteristics that correspond to
surfaces describing waves.


\ack

The author would like to thank the organizers of EmQM13 for their
kind invitation, as well as to Gerhard Gr\"ossing, Johannes Mesa
Pascasio, Herbert Schwabl, and Siegfried Fussy for their hospitality
and interesting discussions during my stay in Vienna.
Profs. Helmut Rauch and Dieter Schuch are also acknowledged for
stimulating discussions during the conference.
Financial support from the Ministerio de Econom{\'\i}a y Competitividad
(Spain) under Project FIS2011-29596-C02-01 and a ``Ram\'on y Cajal''
Research Fellowship, and from the COST Action MP1006 ({\it Fundamental
Problems in Quantum Physics}) is acknowledged.


\section*{References}

\providecommand{\newblock}{}


\begin{thebibliography}{10}
\expandafter\ifx\csname url\endcsname\relax
  \def\url#1{{\tt #1}}\fi
\expandafter\ifx\csname urlprefix\endcsname\relax\def\urlprefix{URL }\fi
\providecommand{\eprint}[2][]{\url{#2}}

\bibitem{bohr100-1:2013}
New Feature 2013 {\em Nature\/} {\bf 498} 21

\bibitem{bohr100-2:2013}
Special Issue 2013 {\em Nature\/} {\bf 498}

\bibitem{young:PTRSL:1804}
Young T 1804 {\em Phil. Trans. R. Soc. Lond.\/} {\bf 94} 1--16

\bibitem{young:1807}
Young T 1807 {\em A Course of Lectures on Natural Philosophy and the Mechanical
  Arts\/} (London: Joseph Johnson)

\bibitem{shamos-bk}
Shamos M~H (ed) 1987 {\em Great Experiments in Physics\/} (New York: Dover)

\bibitem{feynman:FLP3:1965}
Feynman R~P, Leighton R~B and Sands M 1965 {\em The Feynman Lectures on
  Physics\/} vol~3 (Reading, MA: Addison-Wesley)

\bibitem{forman:HistStud:1971}
Forman P 1971 {\em Hist. Stud. Phys. Sci.\/} {\bf 3} 1--115

\bibitem{jammer-bk:1966}
Jammer M 1966 {\em The Conceptual Development of Quantum Mechanics\/} (New
  York: McGraw-Hill)

\bibitem{mermin:PhysToday:1989}
Mermin N~D 1989 {\em Physics Today\/} {\bf 42} 9--11

\bibitem{mermin:PhysToday:2004}
Mermin N~D 2004 {\em Physics Today\/} {\bf 57} 10--11

\bibitem{courant-hilbert-bk-2}
Courant R and Hilbert D 1966 {\em Methods of Mathematical Physics\/} vol~2 (New
  York: John Wiley \& Sons) ch~2

\bibitem{arndt:NNanotech:2012}
Juffmann T, Milic A, M\"ullneritsch M, Asenbaum P, Tsukernik A, T\"uxen J,
  Mayor M, Cheshnovsky O and Arndt M 2012 {\em Nat. Nanotech.\/} {\bf 7}
  297--300

\bibitem{bohm:PR:1952-1}
Bohm D 1952 {\em Phys. Rev.\/} {\bf 85} 166--179

\bibitem{broglie:CompRend-2:1926}
de~Broglie L 1926 {\em Comptes Rendus\/} {\bf 183} 447--448

\bibitem{madelung:ZPhys:1926}
Madelung E 1926 {\em Z. Phys.\/} {\bf 40} 322--326

\bibitem{takabayasi:ProgTheorPhys:1952}
Takabayasi T 1952 {\em Prog. Theor. Phys.\/} {\bf 8} 143--182

\bibitem{takabayasi:ProgTheorPhys:1953}
Takabayasi T 1953 {\em Prog. Theor. Phys.\/} {\bf 9} 187--222

\bibitem{takabayasi:ProgTheorPhys:1983}
Takabayasi T 1983 {\em Prog. Theor. Phys.\/} {\bf 69} 1323--1344

\bibitem{sanz:JPA:2008}
Sanz A~S and Miret-Art\'es S 2008 {\em J. Phys. A: Math. Theor.\/} {\bf 41}
  435303(1--23)

\bibitem{sanz-bk-1}
Sanz A~S and Miret-Art\'es S {\em A Trajectory Description of Quantum
  Processes. I. Fundamentals\/} vol 850 (Berlin: Springer)

\bibitem{kocsis:Science:2011}
Kocsis S, Braverman B, Ravets S, Stevens M~J, Mirin R~P, Shalm L~K and
  Steinberg A~M 2011 {\em Science\/} {\bf 332} 1170--1173

\bibitem{sanz:JPA:2011}
Sanz A~S and Miret-Art\'es S 2011 {\em J. Phys. A: Math. Theor.\/} {\bf 44}
  485301(1--17)

\bibitem{sanz:AnnPhys:2013}
Sanz A~S and Miret-Art\'es S 2013 {\em Ann. Phys.\/} {\bf 339} 11--21
  (\textit{Preprint} \eprint{1104.1296})

\bibitem{mccullough:JCP:1969}
McCullough E~A and Wyatt R~E 1969 {\em J. Chem. Phys.\/} {\bf 51} 1253--1254

\bibitem{mccullough:JCP:1971-1}
McCullough E~A and Wyatt R~E 1971 {\em J. Chem. Phys.\/} {\bf 54} 3578--3591

\bibitem{mccullough:JCP:1971-2}
McCullough E~A and Wyatt R~E 1971 {\em J. Chem. Phys.\/} {\bf 54} 3592--3600

\bibitem{wyatt-bk}
Wyatt R~E 2005 {\em Quantum Dynamics with Trajectories\/} (New York: Springer)

\bibitem{hirsch:JCP:1974-1}
Hirschfelder J~O, Christoph A~C and Palke W~E 1974 {\em J. Chem. Phys.\/} {\bf
  61} 5435--5455

\bibitem{hiley:foundphys:1982}
Dewdney C and Hiley B~J 1982 {\em Found. Phys.\/} {\bf 12} 27--48

\bibitem{lazzeretti:IJQC:1984}
Lazzeretti P, Rossi E and Zanasi R 1984 {\em Int. J. Quantum Chem.\/} {\bf 25}
  929--940

\bibitem{lazzeretti:IJQC:2011}
Pelloni S and Lazzeretti P 2011 {\em Int. J. Quantum Chem.\/} {\bf 111}
  356--367

\bibitem{braunbek:Optik:1952}
Braunbek W and Laukien G 1952 {\em Optik\/} {\bf 9} 174--179

\bibitem{sommerfeld:MathAnn:1896}
Sommerfeld A 1896 {\em Math. Ann.\/} {\bf 47} 317--374

\bibitem{prosser:ijtp:1976-1}
Prosser R~D 1976 {\em Int. J. Theor. Phys.\/} {\bf 15} 169--180

\bibitem{prosser:ijtp:1976-2}
Prosser R~D 1976 {\em Int. J. Theor. Phys.\/} {\bf 15} 181--193

\bibitem{dewdney:NuovoCimB:1979}
Philippidis C, Dewdney C and Hiley B~J 1979 {\em Nuovo Cimento\/} {\bf 52B}
  15--28

\bibitem{sanz:AnnPhysPhoton:2010}
Sanz A~S, Davidovi\'c M, Bo\v{z}i\'c M and Miret-Art\'es S 2010 {\em Ann.
  Phys.\/} {\bf 325} 763--784

\bibitem{zakowicz:PRE:2001}
\.Zakowicz W 2001 {\em Phys. Rev. E\/} {\bf 64} 066610(1--14)

\bibitem{sanz:JOSAA:2012}
Sanz A~S, Campos-Mart{\'\i}nez J and Miret-Art\'es S 2012 {\em J. Opt. Am. Soc.
  A\/} {\bf 29} 695--701

\bibitem{sadreev:PRE:2004}
Sadreev A~F 2004 {\em Phys. Rev. E\/} {\bf 70} 016208(1--7)

\bibitem{scully-zubairy-bk}
Scully M~O and Zubairy S 1997 {\em Quantum Optics\/} (Cambridge: Cambridge
  University Press)

\bibitem{waterhouse:JASA:1985}
Waterhouse R~V, Yates T~W, Feit D and Liu Y~N 1985 {\em J. Acoust. Soc. Am.\/}
  {\bf 78} 758--762

\bibitem{waterhouse:JASA:1986-1}
Waterhouse R~V and Feit D 1986 {\em J. Acoust. Soc. Am.\/} {\bf 80} 681--684

\bibitem{waterhouse:JASA:1986-2}
Skelton E~A and Waterhouse R~V 1986 {\em J. Acoust. Soc. Am.\/} {\bf 80}
  1473--1478

\bibitem{waterhouse:JASA:1987-1}
Waterhouse R~V, Crighton D~G and Ffowcs-Williams J~E 1987 {\em J. Acoust. Soc.
  Am.\/} {\bf 81} 1323--1326

\bibitem{waterhouse:JASA:1987-2}
Waterhouse R~V 1987 {\em J. Acoust. Soc. Am.\/} {\bf 82} 1782--1791

\bibitem{chevalerias:JOSA:1957}
Chevalerias R, Latron Y and Veret C 1957 {\em J. Opt. Soc. Am.\/} {\bf 47}
  703--706

\bibitem{sanz:AJP:2012}
Sanz A~S and Miret-Art\'es S 2012 {\em Am. J. Phys.\/} {\bf 80} 525--533

\bibitem{sanz-bk-2}
Sanz A~S and Miret-Art\'es S {\em A Trajectory Description of Quantum
  Processes. II. Applications\/} vol 831 (Berlin: Springer)

\bibitem{caldirola:NuovoCim:1941}
Caldirola P 1941 {\em Nuovo cim.\/} {\bf 18} 393--400

\bibitem{kanai:ProgTheorPhys:1948}
Kanai E 1948 {\em Prog. Theor. Phys.\/} {\bf 3} 440--442

\bibitem{sanz:arxiv:dissip}
Sanz A~S, Mart{\'\i}nez-Casado R, Pe\~nate-Rodr{\'\i}guez H, Rojas-Lorenzo G
and Miret-Art\'es S 2013 Dissipative Bohmian mechanics: A trajectory analysis of wave-packet dynamics in viscid media
\textit{Preprint} \eprint{1306.6607}

\bibitem{vink:PRA:1993}
Vink J~C 1993 {\em Phys. Rev. A\/} {\bf 48} 1808--1818

\bibitem{lorenzen:PRA:2009}
Lorenzen F, de~Ponte M~A and Moussa M~H~Y 2009 {\em Phys. Rev. A\/} {\bf 80}
  032101(1--8)

\bibitem{hasse:JMathPhys:1975}
Hasse R~W 1975 {\em J. Math. Phys.\/} {\bf 16} 2005--2011

\bibitem{schuch:PRA:1983}
Schuch D, Chung K~M and Hartmann H 1983 {\em Phys. Rev. A\/} {\bf 24}
  1652--1660

\bibitem{schuch:PRA:1984-1}
Schuch D, Chung K~M and Hartmann H 1984 {\em Phys. Rev. A\/} {\bf 25} 391--410

\bibitem{schuch:PRA:1984-2}
Schuch D, Chung K~M and Hartmann H 1984 {\em Phys. Rev. A\/} {\bf 25}
  3086--3092

\bibitem{schuch:PRA:1997}
Schuch D 1997 {\em Phys. Rev. A\/} {\bf 55} 935--940

\bibitem{sanz:CJC:2014}
Sanz A~S 2014 {\em Can. J. Chem.\/} {\bf 92} 168--178

\bibitem{kostin:jcp:1972}
Kostin M~D 1972 {\em J. Chem. Phys.\/} {\bf 57} 3589--3591

\bibitem{garashchuk:JCP:2013}
Garashchuk S, Dixit V, Gu B and Mazzuca J 2013 {\em J. Chem. Phys.\/} {\bf 138}
  054107(1--7)

\bibitem{sanz:PhysScrPhoton:2009}
Davidovi\'c M, Sanz A~S, Arsenovi\'c D, Bo\v{z}i\'c M and Miret-Art\'es S 2009
  {\em Phys. Scr.\/} {\bf T135} 014009(1--5)

\bibitem{bornwolf-bk}
Born M and Wolf E 1999 {\em Principles of Optics. Electromagnetic Theory of
  Propagation, Interference and Diffraction of Light\/} 7th ed (Cambridge:
  Cambridge University Press)

\bibitem{dimic:PhysScr:2013}
Davidovi\'c M, Sanz A~S, Bo\v{z}i\'c M, Arsenovi\'c D and Dimi\'c D 2013 {\em
  Phys. Scr.\/} {\bf T153} 014015(1--5)

\bibitem{feynman-bk2}
Feynman R~P 2006 {\em Path Integral Approach to Quantum Mechanics\/}
  (Sausalito, CA: University Science Books)

\bibitem{barnett:PhilTRSA:2010}
Barnett S~M and Loudon R 2010 {\em Phil. Trans. R. Soc. A\/} {\bf 368} 927--939

\bibitem{sanz:JPCM:2002}
Sanz A~S, Borondo F and Miret-Art\'es S 2002 {\em J. Phys.: Condens. Matter\/}
  {\bf 14} 6109--6145

\end{thebibliography}
\end{document}